\documentclass[sn-nature]{sn-jnl}

\usepackage{graphicx}%
\usepackage{multirow}%
\usepackage{amsmath,amssymb,amsfonts}%
\usepackage{amsthm}%
\usepackage{mathrsfs}%
\usepackage[title]{appendix}%
\usepackage{xcolor}%
\usepackage{textcomp}%
\usepackage{manyfoot}%
\usepackage{booktabs}%
\usepackage{algorithm}%
\usepackage{algorithmicx}%
\usepackage{algpseudocode}%
\usepackage{listings}%
\usepackage{longtable} 
\usepackage{placeins} 

\theoremstyle{thmstyleone}%
\theoremstyle{thmstyletwo}%

\theoremstyle{thmstylethree}%

\begin{document}

\title[Global air quality inequality over 2000-2020]{Global air quality inequality over 2000-2020 \\ {\small July 28, 2023 } }


\author[1]{\fnm{Lutz} \sur{Sager}}\email{lutz.sager@georgetown.edu}

\affil[1]{\orgdiv{McCourt School of Public Policy}, \orgname{Georgetown University}, \orgaddress{\street{3700 O St NW}, \city{Washington}, \postcode{20057}, \state{DC}, \country{United States}}}

\abstract{Air pollution generates substantial health damages and economic costs worldwide \cite{landrigan2018lancet, murray2020global, aguilar2022air, sachs2019six}. Pollution exposure varies greatly, both between countries \cite{southerland2022global, sicard2023trends} and within them \cite{hajat2015socioeconomic, 
jbaily2022air, currie2023caused}. However, the degree of air quality inequality and its' trajectory over time have not been quantified at a global level. Here I use economic inequality indices to measure global inequality in exposure to ambient fine particles with 2.5 microns or less in diameter (PM\textsubscript{2.5}). I find high and rising levels of global air quality inequality. The global PM\textsubscript{2.5} Gini Index increased from 0.32 in 2000 to 0.36 in 2020, exceeding levels of income inequality in many countries. Air quality inequality is mostly driven by differences between countries and less so by variation within them, as decomposition analysis shows. A large share of people facing the highest levels of PM\textsubscript{2.5} exposure are concentrated in only a few countries. The findings suggest that research and policy efforts that focus only on differences within countries are overlooking an important global dimension of environmental justice.}



\maketitle
\vspace{-0.8cm}
Elevated levels of air pollution have been shown to generate vast damages to human health \citep{landrigan2018lancet, murray2020global} and economic productivity \citep{aguilar2022air} worldwide, and improved air quality has been recognized as a key step towards achieving the United Nation's Sustainable Development Goals \cite{sachs2019six}. It is well-established that exposure to ambient air pollution varies substantially within countries, often in ways that correlate with socio-demographic characteristics  \citep{hajat2015socioeconomic, jbaily2022air, currie2023caused}. Research efforts and the public discourse around environmental justice tend to focus on these within-country differences \cite{mohai2009environmental, banzhaf2019environmental, drupp2021inequality}. But pollution exposure also varies greatly between countries \cite{southerland2022global, sicard2023trends}, and tends to be higher in poorer countries \cite{apte2021air, rentschler2023global}. Less is known about the degree of global air quality inequality, and how differences between and within countries shape the global exposure distribution.

In this article, I investigate global air quality inequality between 2000 and 2020 by combining gridded population data \cite{GPWv4} with annual concentration estimates of fine particles with 2.5 microns or less in diameter (PM\textsubscript{2.5}) \cite{van2021monthly}, one of the pollutants most strongly linked to premature deaths and other damages \cite{landrigan2018lancet, murray2020global, aguilar2022air}. The final sample contains 80.1 million grid points in 228 countries or territories\footnote{A full list of countries/territories, following classification in \cite{GPWv4}, is shown in SI Table \ref{CountryList}.} and accounts for 85\% of the world population in 2020 (Extended Data Figures \ref{MapPop2020} and \ref{MapPM2020}).\footnote{Data and code to be be made available online alongside this article upon publication.} 

To quantify global air quality inequality, I rely on indices designed to measure economic inequality, but instead apply them to the global distribution of ambient PM\textsubscript{2.5} exposure. Specifically, I calculate the ratio of the 90th to the 10th percentile of the global PM\textsubscript{2.5} distribution (R9010), the global PM\textsubscript{2.5} Gini Index \cite{gini1921measurement}, and three Generalized Entropy measures including the common Theil Index \cite{shorrocks1980class}. Some of these indices have been used in previous work to characterize the distribution of air pollution within single countries \cite{clark2014national, boyce2016measuring, rosofsky2018temporal, pisoni2022inequality} and to study the distribution of benefits from air quality regulation \cite{levy2007quantifying, fann2011maximizing, holland2019distributional, mansur2021measurement}. Here, I use them to measure air quality inequality at the global level, accounting for variation both between and within countries.

By calculating indices to describe global air quality inequality, I contribute a new perspective to a literature that has thus far focused on describing exposure differences between countries, regions and cities \cite{southerland2022global, sicard2023trends, apte2021air, rentschler2023global}. In particular, the indices allow me to assess whether and by how much global air quality inequality has increased between 2000 and 2020. Some indices allow for sub-group decomposition \cite{shorrocks1984inequality, mookherjee1982decomposition}, which I use to quantify the relative contributions of between-country and within-country differences to global air quality inequality.  

\section*{Results}
\begin{figure}[ht]
\centering
\includegraphics[width=0.5\textwidth]{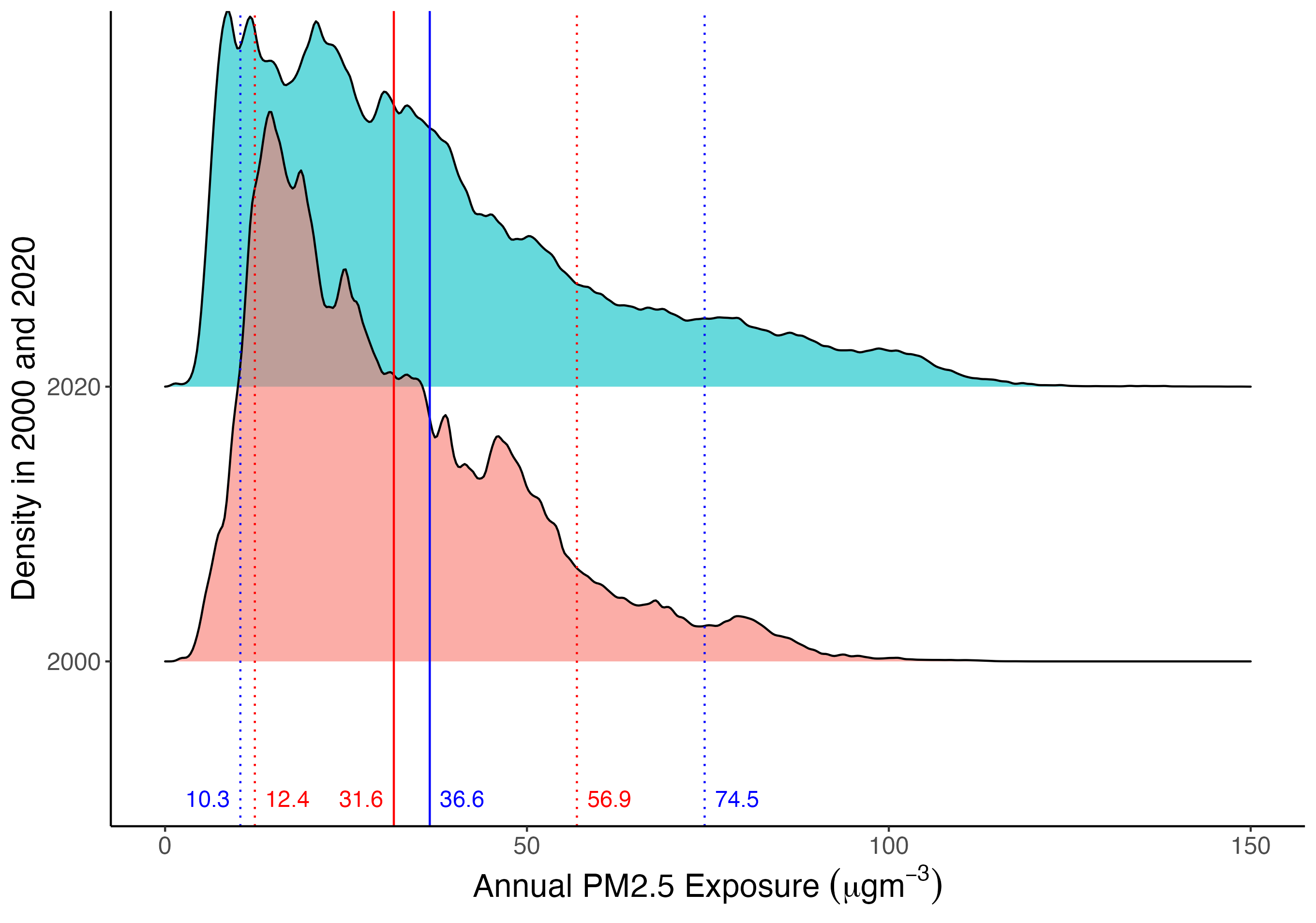}\includegraphics[width=0.5\textwidth]{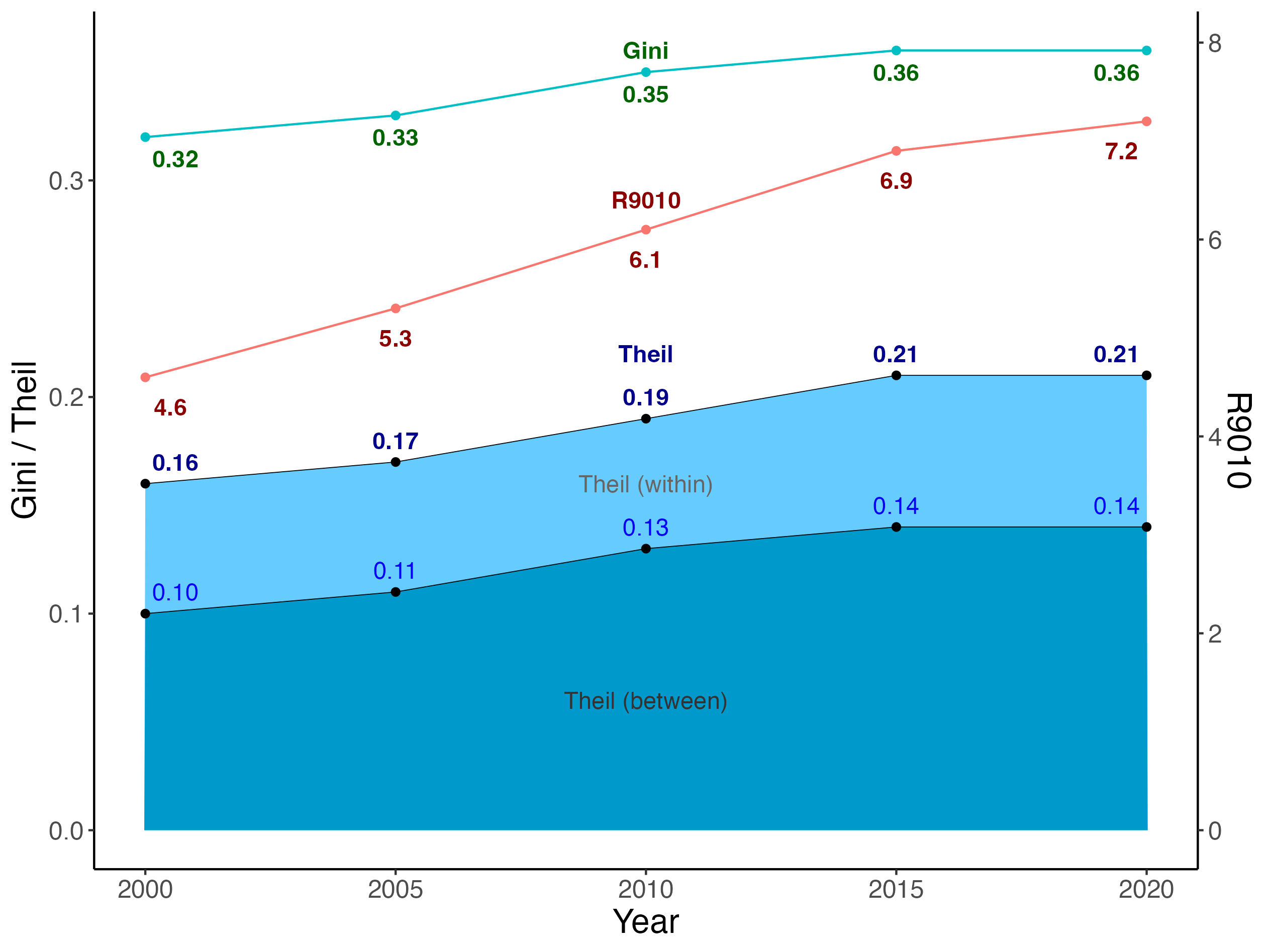} \\
\caption{Global distribution of PM\textsubscript{2.5} exposure between 2000 and 2020. Left panel: Population-weighted Gaussian Kernel density estimates for PM\textsubscript{2.5} levels ($\mu g m^{-3}$) in 2000 and 2020, truncated at 150 $\mu g m^{-3}$. Solid lines indicate global mean PM\textsubscript{2.5} exposure and dashed lines are 10th and 90th percentiles. Right panel: Inequality indices for global PM\textsubscript{2.5} exposure 2000-2020 in 5-year intervals, measured as ratio of 90th to 10th percentile (R9010, right axis), Gini Index and Theil Index (left axis). Theil Index decomposition into between-country and within-country components following \cite{shorrocks1984inequality}. Population and pollution data from \cite{GPWv4, van2021monthly}.}\label{fig:fig1}
\end{figure}

The global distribution of ambient PM\textsubscript{2.5} exposure is plotted in Figure \ref{fig:fig1} (left panel). Population-weighted mean exposure stood at 36.6$\mu g m^{-3}$ in 2020, an increase of 16\% relative to 2000 (31.6$\mu g m^{-3}$). In 2020, over 99\% of the sample population faced PM\textsubscript{2.5} levels exceeding the 5$\mu g m^{-3}$ guideline level set by the World Health Organization in 2021 \cite{world2021global}, and 91\% faced levels exceeding the previous 10$\mu g m^{-3}$ threshold.\footnote{Thresholds in the WHO pollutant guidelines are set based on a systematic review and meta-analysis of the literature linking long-term exposure to PM and all-cause and cause-specific mortality \cite{world2021global}.}

\bmhead{Global air quality inequality}

Exposure to PM\textsubscript{2.5} is highly unequal and has become more unequal over time. Figure \ref{fig:fig1} (left panel) shows that the global PM\textsubscript{2.5} distribution has stretched, with the 90th percentile rising from 58.9$\mu g m^{-3}$ in 2000 to 74.5$\mu g m^{-3}$ in 2020. Meanwhile, the 10th percentile fell from 12.4 to 10.3$\mu g m^{-3}$. Consequently, the ratio of the 90th to the 10th percentile (R9010) increased from 4.6 in 2000 to 7.2 in 2020 (right panel). The least-polluted member of the top decile in 2020 was exposed to over seven times more air particulates than the most polluted member of the bottom decile. The Global Air Quality Gini Index grew from 0.32 in 2000 to 0.36 in 2020, an increase comparable to moving from the degree of income inequality in Germany (0.32 in 2019) to that in India (0.36 in 2019)\cite{worldbank}. The Theil Index, another popular inequality measure, rose from 0.16 in 2000 to 0.21 in 2020, as did the other Generalized Entropy measures (Extended Data Table \ref{Decomposition}).

\bmhead{Decomposition into differences between and within countries}

The bulk of global air quality inequality can be attributed to differences between countries rather than variation within them. Using the Theil Index, which is exactly decomposable \cite{shorrocks1984inequality}, in Figure \ref{fig:fig1} (right panel) we see that two thirds of global air quality inequality (0.14 out of 0.21) were due to differences between countries in 2020. Put differently, even if PM\textsubscript{2.5} exposure was fully equalized within every country in the world, two thirds of global air quality inequality would remain.\footnote{A similar share of global inequality in household carbon footprints inequality (64\% in 2019) is explained by differences between countries.\cite{chancel2022global}} Similar and even higher between-country contributions are shown by the other measures (Extended Data Table \ref{Decomposition}).

\begin{figure}[ht]
\centering
\includegraphics[width=0.5\textwidth]{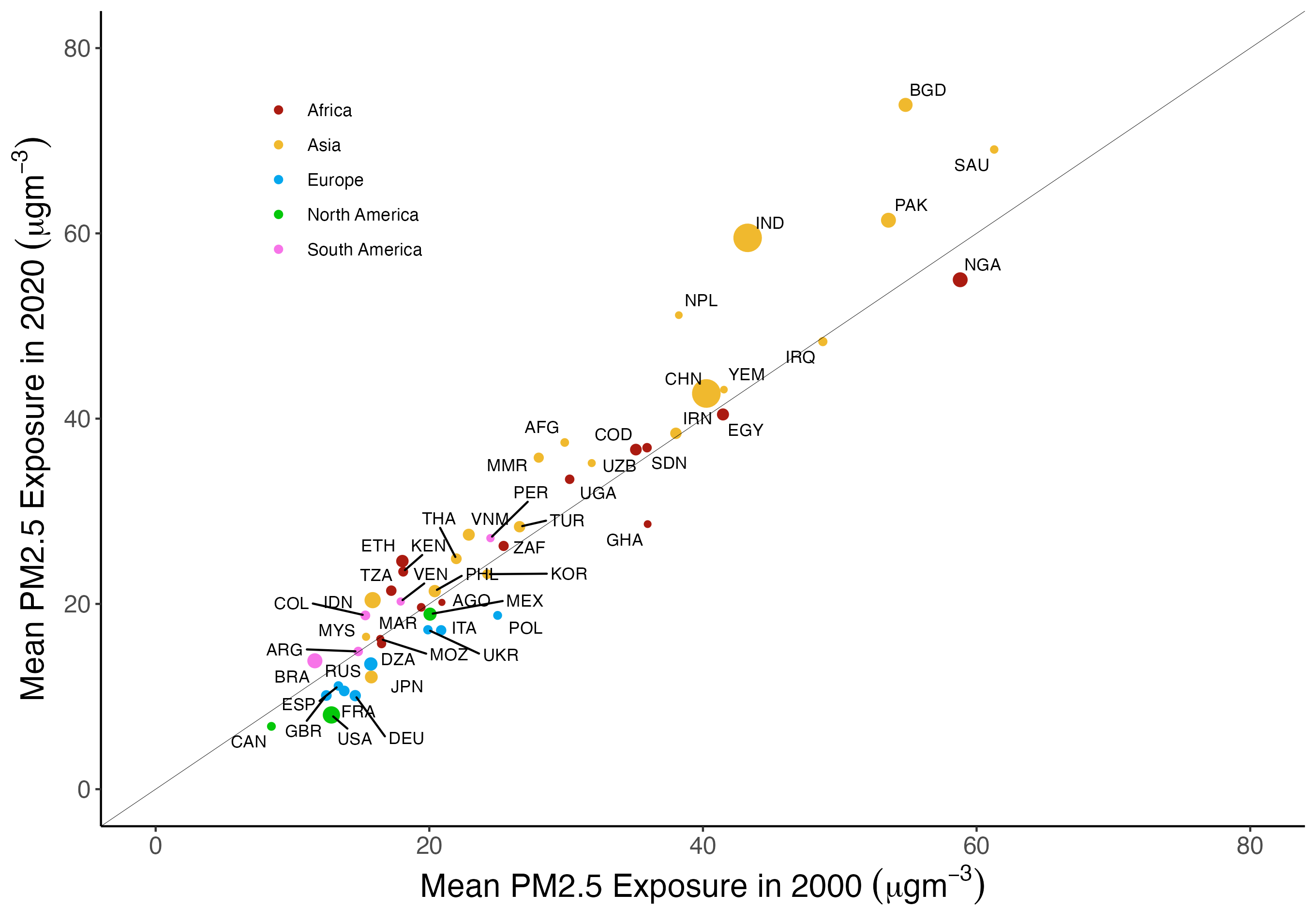}\includegraphics[width=0.5\textwidth]{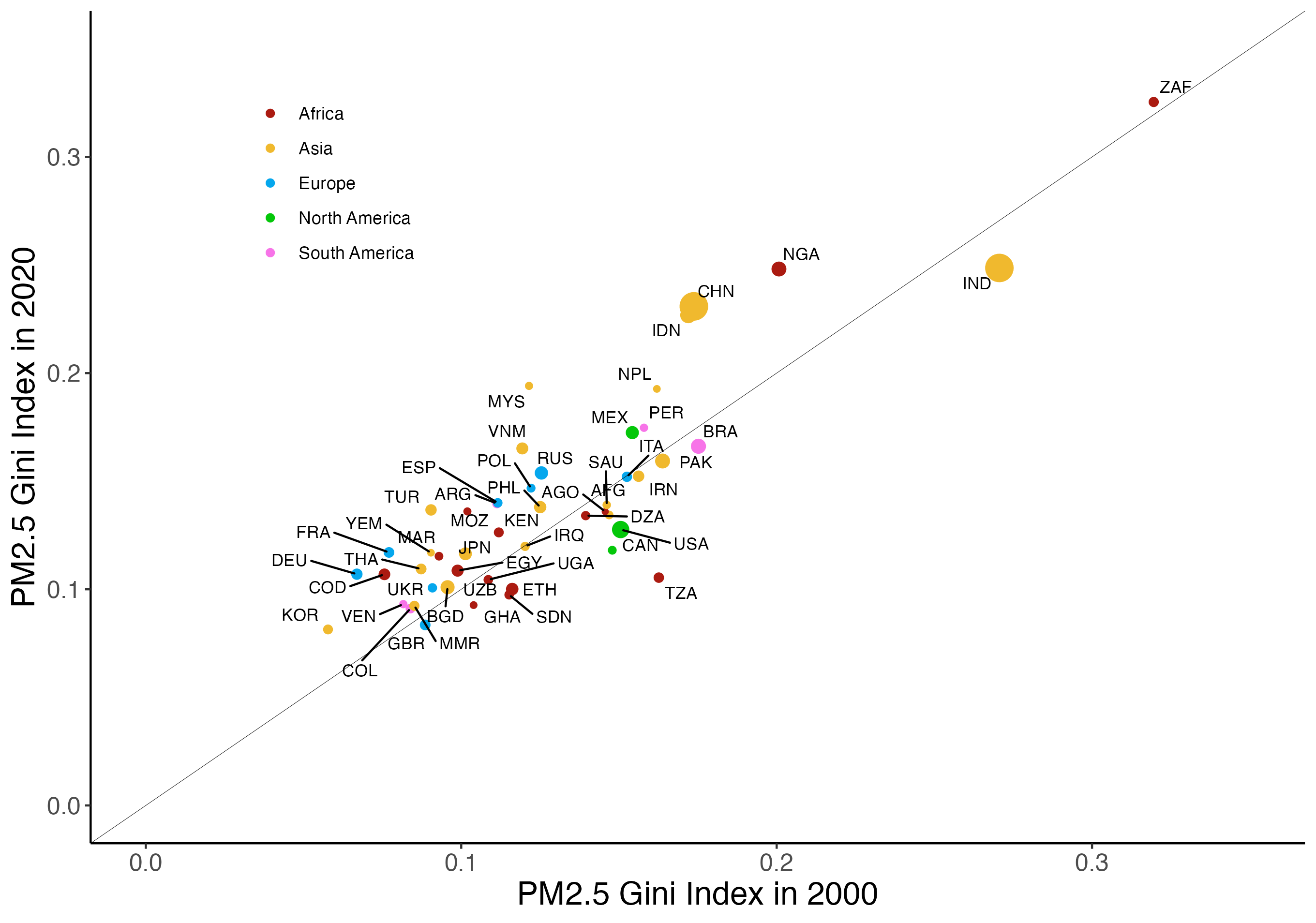} \\
\caption{Country-level PM\textsubscript{2.5} exposure and inequality in 2000 and 2020. Left panel: Country-level population-weighted mean PM\textsubscript{2.5} exposure ($\mu g m^{-3}$) in 2000 and 2020. Right panel: Country-level Gini Index for PM\textsubscript{2.5} exposure in 2000 and 2020. Circle size indicates population in 2020, color indicates continent. Diagonal line represents equal levels in 2000 and 2020. Graph limited to the 50 countries with largest 2020 population. Population and pollution data from \cite{GPWv4, van2021monthly}.}\label{fig:countrypm25}
\end{figure}

Much of the increase in global air quality inequality is accounted for by rising differences between countries. The between-country portion of the Global PM\textsubscript{2.5} Theil Index rose from 0.10 to 0.14 between 2000 and 2020, while the within-country portion grew only slightly from 0.06 to 0.07. Another way to see this is in Figure \ref{fig:countrypm25}, which shows divergence in country-level PM\textsubscript{2.5} exposure between 2000 and 2020 (left panel). Many countries with high PM\textsubscript{2.5} levels in 2000 had even higher levels in 2020. This is visible for South Asian countries such as India, Pakistan and Bangladesh, but also for China, Saudi Arabia and others. Meanwhile, many of the already less polluted countries in Europe and North America saw stagnating or even falling PM\textsubscript{2.5} levels. This reinforced a trend whereby PM\textsubscript{2.5} levels tend to be higher in low-income countries (Extended Data Figure \ref{fig:GDPpc}), likely compounding global economic inequality \cite{apte2021air, rentschler2023global}. Meanwhile, within-country PM\textsubscript{2.5} inequality, measured by the Gini Index (right panel of Figure \ref{fig:countrypm25}), increased in some countries but fell in others. Global PM\textsubscript{2.5} inequality is largely and increasingly driven by differences between countries.

\begin{figure}[h]%
\centering
\includegraphics[width=0.9\textwidth]{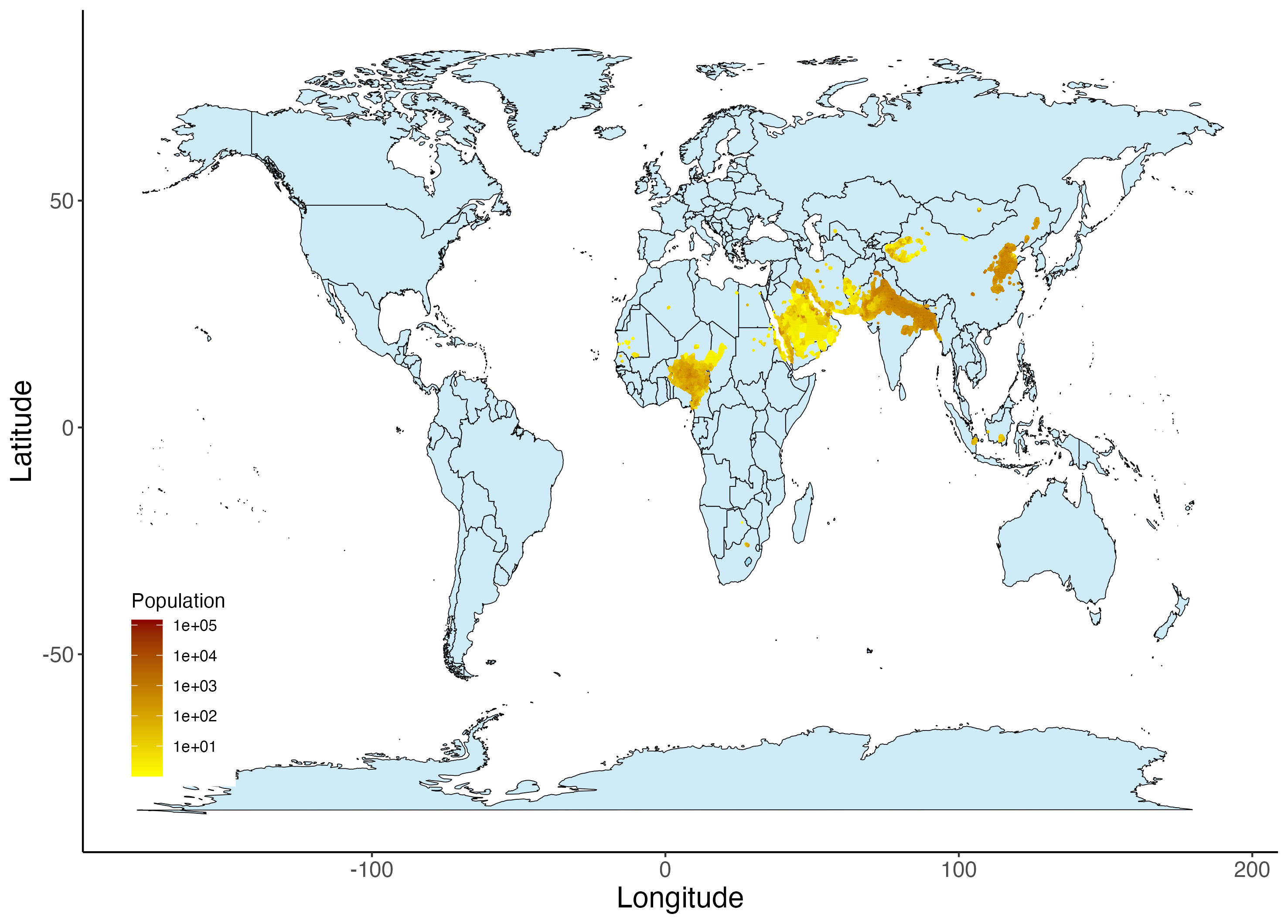}
\caption{The `Choking Billion'. World map showing the location of the 1 billion people exposed to annual PM\textsubscript{2.5} over 63.6 $\mu g m^{-3}$ in 2020. Shading indicates population density. Population and pollution data from \cite{GPWv4, van2021monthly}.} \label{MapChoking}
\end{figure}

\pagebreak
\bmhead{`Air Quality Poverty' and the `Choking Billion'}
High overall PM\textsubscript{2.5} exposure combined with pronounced inequality imply that a substantial portion of the world population faces very high levels of air pollution. Continuing the analogy with economic inequality, we can designate a group of the most exposed people as suffering from `air quality poverty'. Here, I focus on the 1 billion people facing the highest PM\textsubscript{2.5} exposure levels in 2020 (63.6 $\mu g m^{-3}$ or higher), which I call the `Choking Billion'.\footnote{The notion of the `Choking Billion' to describe clusters of extreme air pollution exposure is inspired by the `Bottom Billion' used to describe clusters of economic poverty \cite{collier2008bottom}.} The map in Figure \ref{MapChoking} shows that a few geographic clusters accounted for most of the `Choking Billion' in 2020. The biggest cluster spans Northern India, Bangladesh, Pakistan and Nepal, followed by other big clusters in and around Eastern China, Northern Nigeria, and the Arab peninsula. India alone was home to over half (500 million) of the `Choking Billion' in 2020, followed by China (155m), Bangladesh (127m) and Pakistan (93m), as detailed in Extended Data Table \ref{TableChoking}. Other countries also had high rates of air quality poverty. Notably, two thirds of the population of Saudi Arabia and most inhabitants of Qatar were part of the `Choking Billion' in 2020, as were around half of people in Niger, and almost a third in Nigeria and Mongolia.\footnote{PM\textsubscript{2.5} estimates from \cite{van2021monthly} include mineral dust, which can represent a larger portion of the total count in regions close to deserts.}

\section*{Discussion}\label{Discussion}

The inequality indices calculated here document high and rising levels of global air quality inequality, as measured by annual ambient PM\textsubscript{2.5} exposure. The bulk of this inequality comes from differences between countries rather than variation within them. This suggests that research efforts and the public discourse on environmental justice, which tend to focus on air quality differences within countries \cite{mohai2009environmental, banzhaf2019environmental, drupp2021inequality}, may overlook an important dimension of global environmental inequality. 

Underlying rising mean exposure and exposure inequality is an increase at the upper tails of the PM\textsubscript{2.5} distribution. A large portion of those exposed to the highest levels of ambient PM\textsubscript{2.5} live in just a few countries in Asia and Africa. Much like economic poverty, `air quality poverty' is geographically concentrated, often in places that also face economic hardship \cite{apte2021air, rentschler2023global}. Sustainable development initiatives, which increasingly recognize the importance of air quality improvements \cite{sachs2019six}, may benefit from focusing on these pollution clusters.

The findings are subject to several limitations. Firstly, mismeasurement of either population counts or pollution levels may affect my results, which thus rely on the assumptions underlying the estimates in \cite{GPWv4} and \cite{van2021monthly}. Secondly, the spatial resolution of 30 arc-seconds (500-900 meters depending on latitude) used here necessarily misses variation at smaller geographic scales. Finally, location-based measures of pollution concentrations are imprecise measures of actual exposure. Since PM\textsubscript{2.5} is measured in weight per volume ($\mu g m^{-3}$), assuming that all persons consume similar volumes of outdoor air, the indices capture inequality in total fine particles (by weight) inhaled. Effective exposure will differ based on additional factors, such as mobility patterns, building technology and time spent indoors\cite{jones1999indoor}. 

Despite these limitations, the inequality indices calculated here offer a new, quantitative perspective on global air quality inequality. Similar indices may prove useful in describing other dimensions of environmental inequality. In particular, future research may test if air pollutants other than PM\textsubscript{2.5} are subject to similar inequality levels.

\pagebreak
\pagebreak
\bibliography{Sager_GlobalPM25}

\bigskip
\backmatter

\bmhead{Author contributions}
L.S. conceived and conducted the research, analysed the results, wrote and reviewed the manuscript.

\bmhead{Declarations}
The author declares no competing interests.

\bmhead{Acknowledgments}
I thank Raphael Calel for helpful comments.

\pagebreak

\section*{Methods}\label{MethodsDetail}
This article combines global gridded population counts with estimates of annual fine particulate matter (PM\textsubscript{2.5}) concentrations to calculate inequality indices describing the global distribution of PM\textsubscript{2.5} exposure between 2000 and 2020.

\bmhead{Data} Population count data are from Gridded Population of the World (GPW v4.11) \citep{GPWv4}, using UN WPP-adjusted population counts in 2000, 2005, 2010, 2015 and 2020 at a 30 arc-second (0.0083 degree) resolution. Data on PM\textsubscript{2.5} concentrations are from \cite{van2021monthly} (V5.GL.03), using estimates of annual mean surface-level PM\textsubscript{2.5} concentrations in 2000, 2005, 2010, 2015 and 2020 at a 0.01 arc-degree resolution.

\bmhead{Spatial matching} The spatial unit of analysis are grid cells from GPW v4.11. In each year and for each population grid cell, PM\textsubscript{2.5} is assigned as follows: Where available, assign the PM\textsubscript{2.5} level at the pollution grid point closest (in arc-degrees) to the population grid point. When that PM\textsubscript{2.5} level is missing, use the mean of the non-missing values from the 8 next-closest grid points in arc-degree space (i.e. neighboring grid cells to the top-right, top, top-right, left, right, bottom-left, bottom, and bottom-right). This step extends my sample coverage, but does not substantially change the main results, as shown in SI Table \ref{ADecomposition}.

\bmhead{Final sample} Due to computational constraints, the analysis sample is limited as follows. First, all grid points with population estimates below 1 are dropped (this maintains 99.7\% of the total population). Second, countries/territories with total populations below 10,000 in 2020 are dropped.\footnote{Specifically, this eliminates: Falkland Islands, Montserrat, Niue, Norfolk Islands, Pitcairn, Saint-Barthelemy, Saint Helena, Saint Pierre and Miquelon, Tokelau, Tuvalu.} In addition, the sample excludes grid cells that cannot be matched to a PM\textsubscript{2.5} estimate using the procedure described above. This results in a final sample of 80.1 million grid points in 228 countries/territories, accounting for 85\% of the world population (Extended Data Figures \ref{MapPop2020} and \ref{MapPM2020}).\footnote{The original GPW v4.11 data has a total population of 7,758,982,599 in 2020, 6,584,892,246 of which I successfully match to PM\textsubscript{2.5} levels. Countries and territories are based on the classification in \cite{GPWv4} and population counts are UN WPP-adjusted versions from GWP v4.11. The full list of countries/territories is shown in SI Table \ref{CountryList}.}.

\bmhead{Inequality measures}  
While economic inequality indices are usually applied to the distribution of income or wealth across individuals or households, I apply them to the distribution of ambient PM\textsubscript{2.5} across the world population. In a given year, each geographic grid cell ($i$) is matched to a PM\textsubscript{2.5} concentration ($P_{i}$). Population counts by grid cell ($n_i$) are used as weights to quantify pollution exposure across the world population ($N=\sum_{i} n_i$). Population-weighted global mean PM\textsubscript{2.5} exposure is $\overline{P}=\frac{1}{N}\sum_{i} n_{i}P_{i}$. 

The inequality indices differ in the weight they place on different characteristics of the pollution exposure distribution. R9010 is calculated as the ratio of the 90th percentile of the population-weighted global PM\textsubscript{2.5} distribution divided by the 10th percentile. The Gini Index is defined as a ratio of weighted sums across grid cells $i,j$, so that $\text{Gini} = \frac{\sum_i n_{i} \sum_j n_{j} |P_{i} - P_{j}|}{2\overline{P}N^2}$.

The three remaining measures are members of the Generalized Entropy (GE) class of inequality indices with different values of the inequality aversion parameter $\alpha$. When $\alpha=0$, GE(0), also called the Mean Log Deviation (MLD), is defined as $GE(0) = \frac{1}{N} \sum_{i} \left[ n_i log(\frac{\overline{P}}{P_i}) \right] $. When $\alpha=1$, GE(1), also called the Theil Index, is defined as $GE(1) = \frac{1}{N} \sum_{i} \left[ n_i \frac{P_i}{\overline{P}} log(\frac{P_i}{\overline{P}}) \right] $. And when $\alpha=2$, GE(2), equal to half the square of the coefficient of variation, is defined as $GE(2)=\frac{1}{2} \left( \left[\sum_{i}  \frac{n_i}{N} \frac{P_i}{\overline{P}} \right]^2 -1 \right) $. 

All GE($\alpha$) measures are additively separable and can be decomposed into within-group and between-group inequality \cite{shorrocks1980class, shorrocks1984inequality}, such that $GE(\alpha) = GE_W(\alpha) + GE_B(\alpha)$. Here, inequality within groups ($k$) is defined as $GE_W(\alpha) = \sum_k \left[ (\frac{N_k}{N}) \left(\frac{\overline{P}_k}{\overline{P}} \right)^\alpha GE_k(\alpha) \right]$ where $GE_k(\alpha)$ is calculated as standalone inequality measure for group $k$. Gini Index decomposition follows \cite{mookherjee1982decomposition}, but is not exhaustive when subgroup distributions overlap, resulting in an unexplained residual.

\pagebreak
\begin{appendices}

\section*{Extended Data}\label{secA1}

\begin{figure}[!htb]%
\centering
\includegraphics[width=0.8\textwidth]{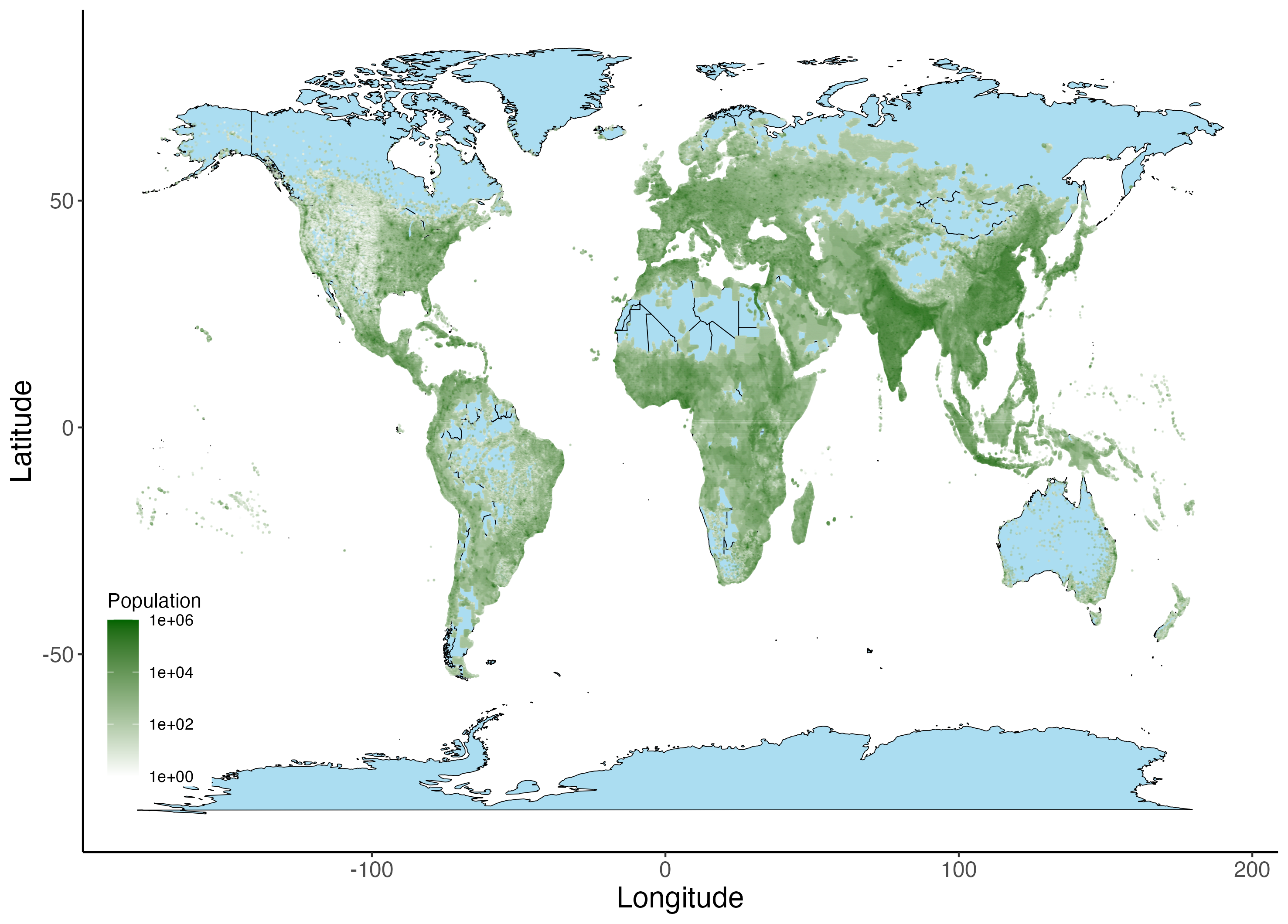}
\caption{Population in 2020. Data from \cite{GPWv4}, limited to analysis sample.} \label{MapPop2020}
\end{figure}

\begin{figure}[!htb]%
\centering
\includegraphics[width=0.8\textwidth]{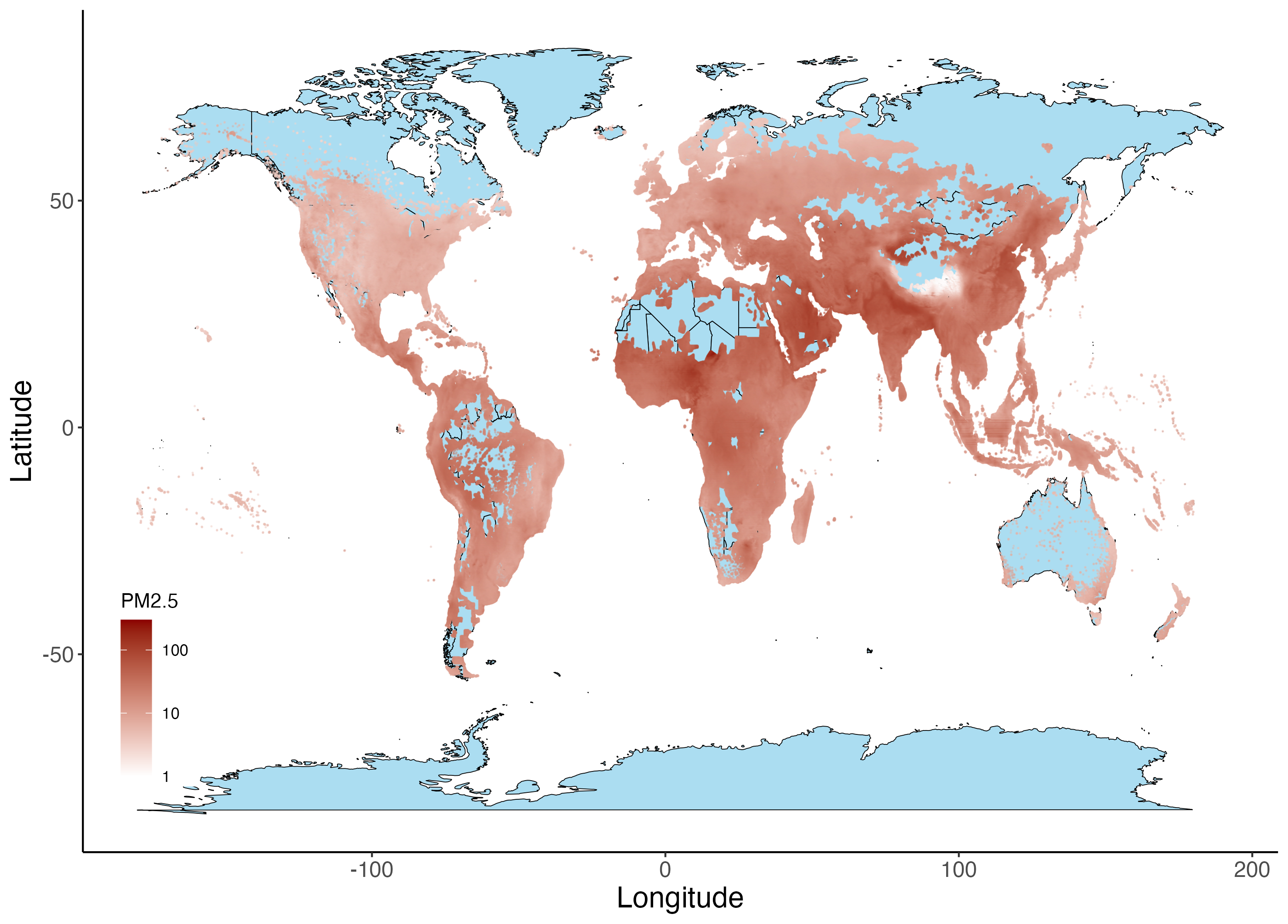}
\caption{PM\textsubscript{2.5} concentrations in 2020. Data from \cite{van2021monthly}, limited to analysis sample.} \label{MapPM2020}
\end{figure}

\begin{figure}[ht]
\centering
\includegraphics[width=0.8\textwidth]{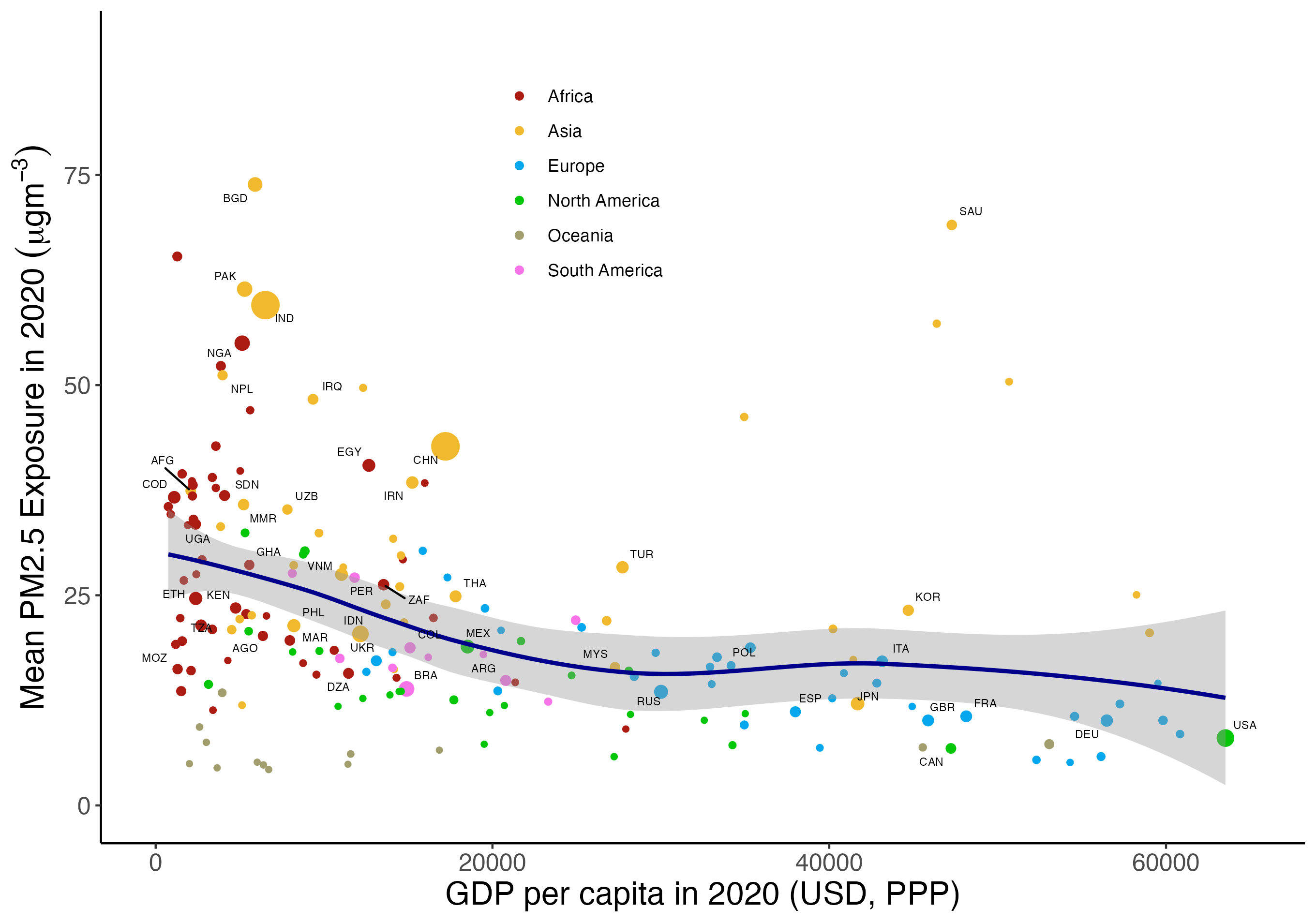} \\
\caption{Country-level GDP per capita and mean PM\textsubscript{2.5} exposure in 2020. Population and pollution data from \cite{GPWv4, van2021monthly}. Plot shows 191 out of 228 countries/regions with available GDP per capita in 2020 (current international dollars, PPP) from \cite{worldbank}. Circle size indicates population in 2020, color indicates continent. Line is fitted from a local polynomial regression with 95\% confidence bands. Labels limited to the 50 countries with largest 2020 population. }\label{fig:GDPpc}
\end{figure}

\begin{table}[h]
\begin{center}
\begin{minipage}{300pt}
\caption{Global PM\textsubscript{2.5} Exposure Inequality and Decomposition}\label{Decomposition}%
\begin{tabular}{@{}lcccccc@{}}
\toprule
 & \hspace*{2mm} Mean \hspace*{1mm} & \hspace*{1mm} R9010 \hspace*{1mm} & \hspace*{1mm} Gini \hspace*{1mm} & \hspace*{1mm} MLD \hspace*{1mm} & \hspace*{1mm} Theil \hspace*{1mm} & \hspace*{1mm} 0.5CV\textsuperscript{2} \hspace*{1mm} \\ 
\midrule
 \textbf{2000} & 31.6 & 4.6 & 0.32 & 0.17 & 0.16 & 0.17  \\ 
 between &  &  & 0.24 & 0.12 & 0.10 & 0.10  \\ 
 within &  &  & 0.02 & 0.05 & 0.06 & 0.07  \\ 
\midrule
 \textbf{2010} & 38.1 & 6.1 & 0.35 & 0.21 & 0.19 & 0.20  \\ 
 between &  &  & 0.27 & 0.16 & 0.13 & 0.12  \\ 
 within &  &  & 0.03 & 0.05 & 0.06 & 0.08  \\ 
\midrule
 \textbf{2020} & 36.6 & 7.2 & 0.36 & 0.23 & 0.21 & 0.22  \\ 
 between &  &  & 0.29 & 0.18 & 0.14 & 0.13  \\ 
 within &  &  & 0.03 & 0.06 & 0.07 & 0.09  \\ 
\botrule
\end{tabular}
\footnotetext{Source: Based on population data from \cite{GPWv4} and PM\textsubscript{2.5} concentrations from \cite{van2021monthly}. `Mean' is population-weighted annual mean PM\textsubscript{2.5} exposure (in $\mu g m^{-3}$); `R9010' is the ratio of the 90th to the 10th percentile; `Gini' is the Gini Index; `MLD' is the mean log deviation or GE(0); `Theil' is the Theil Index or GE(1); 0.5CV\textsuperscript{2} is half of the squared coefficient of variation or GE(2).}
\end{minipage}
\end{center}
\end{table}

\begin{table}[h]
\begin{center}
\begin{minipage}{270pt}
\caption{The `Choking Billion' by country}\label{TableChoking}%
\begin{tabular}{@{}lcccc@{}}
\toprule
 & `Choking'  & `Choking' & Coverage & Total Pop. \\ 
 & (million) & (in \%) & (in \%) & (million) \\ 
\midrule
 India & 509.6 & 40 & 91 & 1262.0 \\ 
 China & 154.8 & 12 & 92 & 1293.2 \\ 
 Bangladesh & 127.2 & 79 & 94 & 160.4 \\ 
 Pakistan & 92.7 & 48 & 93 & 193.5 \\ 
 Nigeria & 46.2 & 29 & 76 & 158.0 \\ 
 Saudi Arabia & 19.5 & 67 & 85 & 29.3 \\ 
 Nepal & 8.7 & 30 & 95 & 28.7 \\ 
 Niger & 6.9 & 46 & 61 & 14.9 \\ 
 Iran & 4.2 & 6 & 82 & 68.5 \\ 
 Cameroon & 3.0 & 18 & 64 & 16.9 \\ 
 Iraq & 2.8 & 8 & 82 & 34.3 \\ 
 United Arab Emirates & 2.1 & 26 & 86 & 8.4 \\ 
 Qatar & 2.0 & 94 & 87 & 2.1 \\ 
 South Africa & 0.9 & 2 & 94 & 53.2 \\ 
 Mongolia & 0.7 & 31 & 69 & 2.2 \\ 
 Kuwait & 0.5 & 14 & 83 & 3.6 \\ 
\botrule
\end{tabular}
\footnotetext{Source: Countries or territories with over 0.5 million residents that are part of the `Choking Billion' in 2020 (facing PM\textsubscript{2.5} levels over 63.6 $\mu g m^{-3}$). Population and pollution data from \cite{GPWv4, van2021monthly}. `Choking' population is rounded to 0.1 million, \% represents share of country population in sample. `Coverage' shows ratio of sample population to total population (`Total Pop.') in \cite{GPWv4}.}
\end{minipage}
\end{center}
\end{table}

\FloatBarrier

\pagebreak
\section*{Supplementary Information}\label{secA2}

\begin{tiny}
\begin{center}
\begin{longtable}{cclcccc}
\caption[xxx.]{Data coverage by country or territory.} \label{CountryList} \\
\toprule
 & ISO & Name of Country or Territory & Pop. in 2020 & Coverage & PM\textsubscript{2.5} in 2000 & PM\textsubscript{2.5} in 2020 \\ 
 &   &  & (thousands) & (\%) & ($\mu g m^{-3}$) & ($\mu g m^{-3}$) \\ 
\midrule
\endfirsthead

\multicolumn{7}{c}%
{{\bfseries \tablename\ \thetable{} -- continued from previous page}} \\
\toprule
 & ISO Code & Name of Country or Territory & Pop. in 2020 & Coverage & PM\textsubscript{2.5} in 2000 & PM\textsubscript{2.5} in 2020 \\ 
 &   &  & (thousands) & (\%) & ($\mu g m^{-3}$) & ($\mu g m^{-3}$) \\ 
\midrule
\endhead

\hline \multicolumn{7}{c}{{Continued on next page}} \\ \hline
\endfoot

\hline \hline
\endlastfoot
1 & ABW & Aruba & 105 & 46.3 & 10.7 & 10.9 \\ 
  2 & AFG & Afghanistan & 36452 & 89.0 & 29.9 & 37.4 \\ 
  3 & AGO & Angola & 29262 & 55.8 & 20.9 & 20.2 \\ 
  4 & AIA & Anguilla & 15 & 81.6 & 11.1 & 9.6 \\ 
  5 & ALA & Åland Islands & 30 & 91.4 & 6.7 & 4.9 \\ 
  6 & ALB & Albania & 2936 & 92.2 & 21.7 & 18.2 \\ 
  7 & AND & Andorra & 69 & 45.2 & 12.1 & 9.0 \\ 
  8 & ARE & United Arab Emirates & 9809 & 85.6 & 61.1 & 58.5 \\ 
  9 & ARG & Argentina & 45530 & 84.3 & 14.8 & 14.9 \\ 
  10 & ARM & Armenia & 3041 & 80.0 & 28.0 & 31.7 \\ 
  11 & ASM & American Samoa & 56 & 91.2 & 3.8 & 4.3 \\ 
  12 & ATG & Antigua and Barbuda & 96 & 79.8 & 13.0 & 11.0 \\ 
  13 & AUS & Australia & 25595 & 90.1 & 5.8 & 7.3 \\ 
  14 & AUT & Austria & 8660 & 71.7 & 16.7 & 12.1 \\ 
  15 & AZE & Azerbaijan & 10232 & 80.4 & 27.8 & 26.0 \\ 
  16 & BDI & Burundi & 13106 & 69.1 & 33.6 & 35.5 \\ 
  17 & BEL & Belgium & 11641 & 90.7 & 14.2 & 10.6 \\ 
  18 & BEN & Benin & 12355 & 77.9 & 43.1 & 39.0 \\ 
  19 & BES & Bonaire Saint Eustatius and Saba & 26 & 76.1 & 12.2 & 11.5 \\ 
  20 & BFA & Burkina Faso & 20865 & 60.5 & 40.7 & 38.1 \\ 
  21 & BGD & Bangladesh & 170397 & 94.1 & 54.8 & 73.9 \\ 
  22 & BGR & Bulgaria & 6909 & 89.7 & 27.7 & 21.2 \\ 
  23 & BHR & Bahrain & 1486 & 88.3 & 52.4 & 50.4 \\ 
  24 & BHS & Bahamas & 410 & 88.8 & 7.7 & 5.8 \\ 
  25 & BIH & Bosnia and Herzegovina & 3757 & 89.9 & 30.3 & 30.3 \\ 
  26 & BLR & Belarus & 9362 & 90.2 & 17.0 & 13.6 \\ 
  27 & BLZ & Belize & 398 & 93.7 & 17.1 & 18.3 \\ 
  28 & BMU & Bermuda & 61 & 94.0 & 7.1 & 5.6 \\ 
  29 & BOL & Bolivia (Plurinational State of) & 11550 & 87.2 & 23.4 & 27.6 \\ 
  30 & BRA & Brazil & 215984 & 76.0 & 11.6 & 13.9 \\ 
  31 & BRB & Barbados & 288 & 59.9 & 14.4 & 13.6 \\ 
  32 & BRN & Brunei Darussalam & 450 & 76.9 & 5.7 & 8.3 \\ 
  33 & BTN & Bhutan & 834 & 94.2 & 23.5 & 28.3 \\ 
  34 & BWA & Botswana & 2458 & 87.2 & 14.9 & 15.2 \\ 
  35 & CAF & Central African Republic & 5399 & 79.2 & 36.3 & 34.6 \\ 
  36 & CAN & Canada & 37599 & 87.9 & 8.5 & 6.8 \\ 
  37 & CHE & Switzerland & 8653 & 74.2 & 14.2 & 10.6 \\ 
  38 & CHL & Chile & 18841 & 89.0 & 19.5 & 22.0 \\ 
  39 & CHN & China & 1402773 & 92.2 & 40.2 & 42.7 \\ 
  40 & CIV & Côte d'Ivoire & 25566 & 93.5 & 23.6 & 22.8 \\ 
  41 & CMR & Cameroon & 26350 & 64.0 & 56.7 & 52.3 \\ 
  42 & COD & Democratic Republic of the Congo & 90176 & 75.4 & 35.1 & 36.7 \\ 
  43 & COG & Congo & 5262 & 66.5 & 28.0 & 37.8 \\ 
  44 & COK & Cook Islands & 21 & 98.6 & 3.9 & 4.2 \\ 
  45 & COL & Colombia & 50230 & 73.2 & 15.3 & 18.8 \\ 
  46 & COM & Comoros & 883 & 66.0 & 10.6 & 11.3 \\ 
  47 & CPV & Cape Verde & 553 & 84.1 & 28.9 & 22.5 \\ 
  48 & CRI & Costa Rica & 5045 & 71.7 & 16.5 & 19.5 \\ 
  49 & CUB & Cuba & 11366 & 93.5 & 11.5 & 10.4 \\ 
  50 & CUW & Curaçao & 164 & 73.6 & 11.5 & 11.9 \\ 
  51 & CYM & Cayman Islands & 64 & 98.2 & 10.9 & 11.1 \\ 
  52 & CYP & Cyprus & 1218 & 80.3 & 20.3 & 17.3 \\ 
  53 & CZE & Czech Republic & 10576 & 73.0 & 20.7 & 14.6 \\ 
  54 & DEU & Germany & 80389 & 69.9 & 14.6 & 10.1 \\ 
  55 & DJI & Djibouti & 946 & 66.8 & 39.4 & 39.8 \\ 
  56 & DMA & Dominica & 74 & 61.5 & 11.7 & 11.8 \\ 
  57 & DNK & Denmark & 5776 & 61.0 & 13.1 & 8.5 \\ 
  58 & DOM & Dominican Republic & 11112 & 95.1 & 11.8 & 12.6 \\ 
  59 & DZA & Algeria & 43007 & 82.7 & 16.5 & 15.7 \\ 
  60 & ECU & Ecuador & 17340 & 70.3 & 14.0 & 17.5 \\ 
  61 & EGY & Egypt & 100524 & 94.0 & 41.5 & 40.5 \\ 
  62 & ERI & Eritrea & 5893 & 68.9 & 32.5 & 32.7 \\ 
  63 & ESH & Western Sahara & 634 & 70.2 & 35.6 & 29.7 \\ 
  64 & ESP & Spain & 46178 & 84.2 & 13.4 & 11.1 \\ 
  65 & EST & Estonia & 1293 & 88.7 & 8.8 & 6.9 \\ 
  66 & ETH & Ethiopia & 111983 & 65.3 & 18.0 & 24.6 \\ 
  67 & FIN & Finland & 5555 & 86.8 & 6.9 & 5.4 \\ 
  68 & FJI & Fiji & 915 & 98.9 & 6.5 & 6.1 \\ 
  69 & FRA & France & 65734 & 84.2 & 13.8 & 10.6 \\ 
  70 & FRO & Faeroe Islands & 49 & 90.0 & 5.2 & 5.4 \\ 
  71 & FSM & Micronesia (Federated States of) & 108 & 83.9 & 4.2 & 4.5 \\ 
  72 & GAB & Gabon & 1919 & 56.7 & 23.8 & 29.3 \\ 
  73 & GBR & United Kingdom & 66699 & 78.8 & 12.5 & 10.1 \\ 
  74 & GEO & Georgia & 3980 & 80.0 & 20.0 & 21.8 \\ 
  75 & GGY & Guernsey & 63 & 78.0 & 10.4 & 8.9 \\ 
  76 & GHA & Ghana & 30548 & 79.8 & 36.0 & 28.6 \\ 
  77 & GIB & Gibraltar & 34 & 84.0 & 13.3 & 13.0 \\ 
  78 & GIN & Guinea & 14354 & 50.9 & 31.6 & 29.2 \\ 
  79 & GLP & Guadeloupe & 419 & 81.5 & 11.6 & 11.6 \\ 
  80 & GMB & Gambia & 2319 & 63.0 & 47.6 & 38.6 \\ 
  81 & GNB & Guinea-Bissau & 2070 & 52.9 & 39.9 & 33.3 \\ 
  82 & GNQ & Equatorial Guinea & 970 & 53.5 & 35.6 & 38.4 \\ 
  83 & GRC & Greece & 10828 & 87.9 & 19.6 & 15.3 \\ 
  84 & GRD & Grenada & 109 & 64.6 & 14.4 & 13.1 \\ 
  85 & GRL & Greenland & 56 & 60.4 & 2.1 & 1.7 \\ 
  86 & GTM & Guatemala & 18015 & 71.6 & 31.2 & 30.3 \\ 
  87 & GUF & French Guiana & 305 & 63.8 & 14.6 & 16.0 \\ 
  88 & GUM & Guam & 180 & 64.7 & 3.2 & 4.4 \\ 
  89 & GUY & Guyana & 787 & 70.0 & 17.2 & 18.0 \\ 
  90 & HKG & Hong Kong & 7619 & 89.9 & 25.6 & 20.5 \\ 
  91 & HND & Honduras & 8656 & 68.9 & 28.6 & 32.4 \\ 
  92 & HRV & Croatia & 4164 & 80.5 & 21.2 & 18.2 \\ 
  93 & HTI & Haiti & 11373 & 93.9 & 12.7 & 14.4 \\ 
  94 & HUN & Hungary & 9684 & 89.6 & 23.0 & 16.6 \\ 
  95 & IDN & Indonesia & 271854 & 76.1 & 15.9 & 20.4 \\ 
  96 & IMN & Isle of Man & 91 & 78.5 & 8.6 & 8.1 \\ 
  97 & IND & India & 1388953 & 90.9 & 43.3 & 59.5 \\ 
  98 & IRL & Ireland & 4875 & 81.2 & 9.4 & 8.1 \\ 
  99 & IRN & Iran (Islamic Republic of) & 83401 & 82.1 & 38.0 & 38.4 \\ 
  100 & IRQ & Iraq & 41963 & 81.7 & 48.8 & 48.3 \\ 
  101 & ISL & Iceland & 342 & 81.3 & 5.2 & 5.1 \\ 
  102 & ISR & Israel & 8733 & 83.5 & 20.1 & 21.0 \\ 
  103 & ITA & Italy & 59743 & 66.3 & 20.9 & 17.2 \\ 
  104 & JAM & Jamaica & 2840 & 93.9 & 16.7 & 18.4 \\ 
  105 & JEY & Jersey & 105 & 88.9 & 10.8 & 9.3 \\ 
  106 & JOR & Jordan & 8167 & 83.9 & 28.1 & 32.4 \\ 
  107 & JPN & Japan & 125039 & 89.7 & 15.8 & 12.1 \\ 
  108 & KAZ & Kazakhstan & 18626 & 82.0 & 17.3 & 22.0 \\ 
  109 & KEN & Kenya & 52175 & 71.3 & 18.1 & 23.5 \\ 
  110 & KGZ & Kyrgyzstan & 6426 & 91.2 & 22.2 & 22.2 \\ 
  111 & KHM & Cambodia & 16809 & 67.4 & 18.5 & 20.9 \\ 
  112 & KIR & Kiribati & 122 & 39.8 & 4.3 & 5.0 \\ 
  113 & KNA & Saint Kitts and Nevis & 58 & 80.3 & 12.6 & 10.8 \\ 
  114 & KOR & Republic of Korea & 51252 & 89.6 & 24.2 & 23.2 \\ 
  115 & KWT & Kuwait & 4314 & 83.0 & 59.6 & 57.3 \\ 
  116 & LAO & Lao People's Democratic Republic & 7407 & 89.8 & 23.2 & 28.6 \\ 
  117 & LBN & Lebanon & 5897 & 84.4 & 28.4 & 29.7 \\ 
  118 & LBR & Liberia & 5093 & 69.3 & 20.1 & 22.3 \\ 
  119 & LBY & Libya & 6700 & 67.8 & 26.2 & 22.3 \\ 
  120 & LCA & Saint Lucia & 192 & 62.8 & 13.8 & 12.7 \\ 
  121 & LIE & Liechtenstein & 39 & 47.5 & 14.2 & 10.5 \\ 
  122 & LKA & Sri Lanka & 21157 & 76.7 & 19.6 & 23.9 \\ 
  123 & LSO & Lesotho & 2241 & 95.9 & 24.6 & 27.5 \\ 
  124 & LTU & Lithuania & 2795 & 89.2 & 15.6 & 12.7 \\ 
  125 & LUX & Luxembourg & 599 & 92.3 & 13.3 & 9.8 \\ 
  126 & LVA & Latvia & 1920 & 88.2 & 17.5 & 14.4 \\ 
  127 & MAC & Macau & 632 & 99.9 & 29.5 & 25.1 \\ 
  128 & MAF & Saint-Martin (French part) & 46 & 69.8 & 10.7 & 9.3 \\ 
  129 & MAR & Morocco & 36457 & 85.7 & 19.4 & 19.6 \\ 
  130 & MCO & Monaco & 26 & 75.9 & 16.7 & 12.5 \\ 
  131 & MDA & Republic of Moldova & 4015 & 91.1 & 18.2 & 15.9 \\ 
  132 & MDG & Madagascar & 27799 & 82.9 & 11.3 & 13.6 \\ 
  133 & MDV & Maldives & 393 & 56.8 & 15.3 & 16.2 \\ 
  134 & MEX & Mexico & 134787 & 93.7 & 20.1 & 18.9 \\ 
  135 & MHL & Marshall Islands & 53 & 78.0 & 4.3 & 5.1 \\ 
  136 & MKD & Macedonia & 2087 & 90.0 & 32.3 & 27.1 \\ 
  137 & MLI & Mali & 20458 & 65.2 & 38.7 & 34.0 \\ 
  138 & MLT & Malta & 423 & 64.2 & 13.0 & 11.8 \\ 
  139 & MMR & Myanmar & 56256 & 93.1 & 28.0 & 35.8 \\ 
  140 & MNE & Montenegro & 626 & 90.1 & 21.2 & 20.8 \\ 
  141 & MNG & Mongolia & 3180 & 68.9 & 32.2 & 49.7 \\ 
  142 & MNP & Northern Mariana Islands & 56 & 91.6 & 2.9 & 3.4 \\ 
  143 & MOZ & Mozambique & 32025 & 75.6 & 16.4 & 16.2 \\ 
  144 & MRT & Mauritania & 4569 & 63.1 & 55.1 & 47.0 \\ 
  145 & MTQ & Martinique & 395 & 62.9 & 16.7 & 16.3 \\ 
  146 & MUS & Mauritius & 1291 & 80.9 & 15.5 & 14.6 \\ 
  147 & MWI & Malawi & 19993 & 65.6 & 16.4 & 19.6 \\ 
  148 & MYS & Malaysia & 32374 & 68.0 & 15.4 & 16.5 \\ 
  149 & MYT & Mayotte & 273 & 62.7 & 12.3 & 13.0 \\ 
  150 & NAM & Namibia & 2722 & 77.4 & 15.6 & 15.6 \\ 
  151 & NCL & New Caledonia & 280 & 98.1 & 6.3 & 6.2 \\ 
  152 & NER & Niger & 24316 & 61.4 & 66.4 & 65.3 \\ 
  153 & NGA & Nigeria & 206824 & 76.4 & 58.8 & 55.0 \\ 
  154 & NIC & Nicaragua & 6416 & 70.5 & 18.2 & 20.7 \\ 
  155 & NLD & Netherlands & 17184 & 89.9 & 14.8 & 10.1 \\ 
  156 & NOR & Norway & 5490 & 65.2 & 8.1 & 5.7 \\ 
  157 & NPL & Nepal & 30197 & 95.1 & 38.2 & 51.2 \\ 
  158 & NRU & Nauru & 10 & 100.0 & 4.6 & 4.9 \\ 
  159 & NZL & New Zealand & 4730 & 88.8 & 6.3 & 6.9 \\ 
  160 & OMN & Oman & 4826 & 85.1 & 47.6 & 46.2 \\ 
  161 & PAK & Pakistan & 208366 & 92.9 & 53.6 & 61.4 \\ 
  162 & PAN & Panama & 4230 & 69.4 & 13.7 & 16.0 \\ 
  163 & PER & Peru & 33317 & 72.6 & 24.5 & 27.1 \\ 
  164 & PHL & Philippines & 108436 & 72.7 & 20.4 & 21.4 \\ 
  165 & PLW & Palau & 22 & 90.3 & 6.6 & 6.6 \\ 
  166 & PNG & Papua New Guinea & 8413 & 89.1 & 9.7 & 13.4 \\ 
  167 & POL & Poland & 38409 & 88.3 & 25.0 & 18.8 \\ 
  168 & PRI & Puerto Rico & 3675 & 94.8 & 8.1 & 7.2 \\ 
  169 & PRK & Democratic People's Republic of Korea & 25760 & 89.8 & 23.8 & 25.6 \\ 
  170 & PRT & Portugal & 10163 & 67.1 & 11.7 & 9.6 \\ 
  171 & PRY & Paraguay & 7064 & 84.8 & 14.0 & 16.4 \\ 
  172 & PSE & State of Palestine & 5317 & 83.9 & 21.2 & 22.6 \\ 
  173 & PYF & French Polynesia & 296 & 97.0 & 4.5 & 4.8 \\ 
  174 & QAT & Qatar & 2452 & 87.1 & 76.5 & 87.0 \\ 
  175 & REU & Réunion & 892 & 83.4 & 6.6 & 6.1 \\ 
  176 & ROU & Romania & 18829 & 88.4 & 20.9 & 17.6 \\ 
  177 & RUS & Russian Federation & 142900 & 80.3 & 15.7 & 13.5 \\ 
  178 & RWA & Rwanda & 13013 & 53.9 & 36.4 & 36.8 \\ 
  179 & SAU & Saudi Arabia & 34371 & 85.1 & 61.3 & 69.1 \\ 
  180 & SDN & Sudan & 45307 & 70.2 & 35.9 & 36.9 \\ 
  181 & SEN & Senegal & 17493 & 59.8 & 49.1 & 42.7 \\ 
  182 & SGP & Singapore & 6007 & 44.8 & 17.3 & 16.5 \\ 
  183 & SLB & Solomon Islands & 640 & 78.1 & 6.5 & 9.3 \\ 
  184 & SLE & Sierra Leone & 7160 & 55.1 & 26.1 & 26.8 \\ 
  185 & SLV & El Salvador & 6227 & 70.1 & 28.8 & 29.9 \\ 
  186 & SMR & San Marino & 31 & 58.2 & 17.8 & 14.5 \\ 
  187 & SOM & Somalia & 12436 & 62.3 & 20.2 & 19.2 \\ 
  188 & SRB & Serbia & 6645 & 90.2 & 23.0 & 23.4 \\ 
  189 & SSD & South Sudan & 14116 & 71.9 & 22.7 & 22.2 \\ 
  190 & STP & Sao Tome and Principe & 211 & 97.5 & 20.7 & 17.2 \\ 
  191 & SUR & Suriname & 565 & 74.9 & 17.1 & 17.6 \\ 
  192 & SVK & Slovakia & 5434 & 89.3 & 22.3 & 16.5 \\ 
  193 & SVN & Slovenia & 2077 & 65.1 & 20.2 & 15.7 \\ 
  194 & SWE & Sweden & 10121 & 74.5 & 8.4 & 5.8 \\ 
  195 & SWZ & Swaziland & 1366 & 95.8 & 21.5 & 16.9 \\ 
  196 & SXM & Sint Maarten (Dutch part) & 42 & 83.1 & 11.1 & 10.1 \\ 
  197 & SYC & Seychelles & 99 & 67.6 & 9.2 & 9.1 \\ 
  198 & SYR & Syrian Arab Republic & 20987 & 81.2 & 29.7 & 32.6 \\ 
  199 & TCA & Turks and Caicos Islands & 37 & 94.0 & 9.3 & 7.3 \\ 
  200 & TCD & Chad & 16422 & 61.6 & 40.3 & 39.4 \\ 
  201 & TGO & Togo & 8273 & 63.5 & 43.4 & 33.5 \\ 
  202 & THA & Thailand & 68596 & 76.4 & 22.0 & 24.9 \\ 
  203 & TJK & Tajikistan & 9416 & 91.0 & 25.5 & 33.2 \\ 
  204 & TKM & Turkmenistan & 5702 & 80.0 & 36.7 & 31.3 \\ 
  205 & TLS & Timor-Leste & 1317 & 64.7 & 11.1 & 11.9 \\ 
  206 & TON & Tonga & 111 & 99.1 & 4.8 & 4.3 \\ 
  207 & TTO & Trinidad and Tobago & 1378 & 61.0 & 15.4 & 15.5 \\ 
  208 & TUN & Tunisia & 11836 & 62.2 & 23.0 & 18.5 \\ 
  209 & TUR & Turkey & 82262 & 86.1 & 26.6 & 28.3 \\ 
  210 & TWN & Taiwan & 23402 & 93.5 & 24.2 & 19.8 \\ 
  211 & TZA & United Republic of Tanzania & 62280 & 65.4 & 17.2 & 21.4 \\ 
  212 & UGA & Uganda & 45836 & 73.8 & 30.3 & 33.5 \\ 
  213 & UKR & Ukraine & 43678 & 85.5 & 19.9 & 17.2 \\ 
  214 & URY & Uruguay & 3492 & 82.1 & 12.0 & 12.3 \\ 
  215 & USA & United States of America & 333422 & 90.3 & 12.8 & 8.0 \\ 
  216 & UZB & Uzbekistan & 31709 & 88.4 & 31.9 & 35.2 \\ 
  217 & VCT & Saint Vincent and the Grenadines & 111 & 76.2 & 14.3 & 13.6 \\ 
  218 & VEN & Venezuela (Bolivarian Republic of) & 33118 & 69.0 & 17.9 & 20.3 \\ 
  219 & VGB & British Virgin Islands & 33 & 94.6 & 7.7 & 7.2 \\ 
  220 & VIR & United States Virgin Islands & 107 & 91.8 & 7.0 & 6.5 \\ 
  221 & VNM & Viet Nam & 98139 & 80.5 & 22.9 & 27.5 \\ 
  222 & VUT & Vanuatu & 294 & 87.7 & 7.5 & 7.5 \\ 
  223 & WLF & Wallis and Futuna Islands & 13 & 77.8 & 4.4 & 4.7 \\ 
  224 & WSM & Western Samoa & 199 & 70.8 & 4.5 & 4.8 \\ 
  225 & YEM & Yemen & 30029 & 63.9 & 41.5 & 43.1 \\ 
  226 & ZAF & South Africa & 56689 & 93.9 & 25.4 & 26.3 \\ 
  227 & ZMB & Zambia & 18889 & 72.3 & 18.5 & 20.9 \\ 
  228 & ZWE & Zimbabwe & 17464 & 93.8 & 15.3 & 16.0 \\ 
\hline
\multicolumn{7}{l}{Source: The 228 countries or territories in the sample, sorted by 2020 population. Names and population estimates from \cite{GPWv4}.} \\
\multicolumn{7}{l}{`Coverage' shows ratio of country population in sample to total country population.} \\
\multicolumn{7}{l}{`PM\textsubscript{2.5} in 2000/2020' are population-weighted mean exposure levels in 2000 and 2020 respectively, based on \cite{van2021monthly}.} \\
\end{longtable}
\end{center}
\end{tiny}

\pagebreak
\begin{table}[h]
\begin{center}
\begin{minipage}{300pt}
\caption{Replication of Table 1 (sample restricted to nearest neighbor PM\textsubscript{2.5})}\label{ADecomposition}%
\begin{tabular}{@{}lcccccc@{}}
\toprule
 & \hspace*{2mm} Mean \hspace*{1mm} & \hspace*{1mm} R9010 \hspace*{1mm} & \hspace*{1mm} Gini \hspace*{1mm} & \hspace*{1mm} MLD \hspace*{1mm} & \hspace*{1mm} Theil \hspace*{1mm} & \hspace*{1mm} 0.5CV\textsuperscript{2} \hspace*{1mm} \\ 
\midrule
   \textbf{2000} & 31.10 & 4.60 & 0.32 & 0.17 & 0.16 & 0.17 \\ 
    between &  &  & 0.25 & 0.12 & 0.11 & 0.10 \\ 
    within &  &  & 0.02 & 0.05 & 0.06 & 0.07 \\ 
   \midrule
    \textbf{2010} & 37.30 & 6.00 & 0.35 & 0.21 & 0.19 & 0.20 \\ 
    between &  &  & 0.28 & 0.16 & 0.13 & 0.12 \\ 
    within &  &  & 0.02 & 0.05 & 0.06 & 0.08 \\ 
   \midrule
    \textbf{2020} & 34.10 & 7.20 & 0.37 & 0.24 & 0.22 & 0.24 \\ 
    between &  &  & 0.31 & 0.18 & 0.16 & 0.16 \\ 
    within &  &  & 0.02 & 0.05 & 0.06 & 0.09 \\ 
\botrule
\end{tabular}
\footnotetext{Source: Replication of Extended Data Table 1, using restricted sample of grid cells assigned non-missing PM\textsubscript{2.5} readings from only the single closest pollution grid cell (44.5m cells accounting for 46\% of world population in 2020). Based on population data from \cite{GPWv4} and PM\textsubscript{2.5} concentrations from \cite{van2021monthly}. `Mean' is population-weighted annual mean PM\textsubscript{2.5} exposure (in $\mu g m^{-3}$); `R9010' is the ratio of the 90th to the 10th percentile; `Gini' is the Gini Index; `MLD' is the mean log deviation or GE(0); `Theil' is the Theil Index or GE(1); 0.5CV\textsuperscript{2} is half of the squared coefficient of variation or GE(2).}
\end{minipage}
\end{center}
\end{table}




\end{appendices}

\end{document}